\documentstyle[aps,preprint,eqsecnum]{revtex}
\begin{document}
\draft
\preprint{October 25, 1995}
\title{
Hidden Massive Spectators in the Effective Field Theory\\
for Integral Quantum Hall Transitions}
\author{Yasuhiro Hatsugai$^{1*}$, Mahito Kohmoto$^2$,
and Yong-Shi Wu$^3$}
\address{
$^1$ Department of Applied Physics,
 University of Tokyo,
 7-3-1 Hongo Bunkyo-ku, Tokyo 113, Japan \\
$^2$ Institute for Solid State Physics,
 University of Tokyo,
 7-22-1 Roppongi Minato-ku,\\
 Tokyo 106, Japan \\
$^3$ Department of Physics, University of Utah,
 Salt Lake City, Utah 84112, U.S.A.
}
\date{October 25, 1995}
\maketitle

\begin{abstract}
Integral quantum Hall plateau transitions in a planar
lattice system due to gap collapse
can be described by an effective
field theory with Dirac fermions. We discuss
how to reproduce the correct integral values for the
Hall conductance, $\sigma_{xy}$,
before and after the plateau
transition, which are dictated by the microscopic
topological invariant. In addition to the massless Dirac
fermions that appear at gap closing, the
matching condition of $\sigma_{xy}$
requires the introduction of massive
Dirac fermions, as ``spectators'', in the effective field
theory.
For non-interacting electrons  on the lattice,
we give a general prescription to determine
these massive ``spectators'', based on microscopic
information  on the vortices in the magnetic Brillouin
zone which are closely related to edge states.
Our description is demonstrated  in a model
with both nearest-neighbor and next-nearest-neighbor hoppings.
\end{abstract}

\section{INTRODUCTION}
\label{sec:intro}
Integral quantum Hall effect in a planar electron
system is a beautiful manifestation of topological
effects in physics. Whenever the Fermi level is in
an energy gap, the quantized Hall conductance is known
to be a topological invariant at the microscopic level.
 \cite{tknn,mkann,ntw,avron1,avron2,ish,edge_yh,chern_yh}
For example, in a periodic system the Hall conductance
for a filled  band is the Chern number over the magnetic
Brillouin zone which is topologically a
torus, \cite{tknn,mkann} usually called the TKNN invariant
named after its discovers. For a lattice system with edges,
the Hall conductance can be written as another topological
invariant for the complex energy surface which is normally
a higher-genus Riemann surface. \cite{edge_yh} (There is a
close relationship between this invariant and edge
states \cite{chern_yh}, which we are going to exploit
in this paper.)
Thus the quantization of Hall conductance is expected to
be stable against weak disorder and perturbations unless
some topological change, e.g. gap closing, occurs.
 \cite{gap_k,gap,wz}
When the gap is reopened, usually there is a discrete change
in the Hall conductance. Such a Hall plateau transition
due to gap collapse is an example of the so-called quantum phase
transitions, which recently attracts some theoretical
attention. \cite{ww,cfw,zhang}

For definiteness, let us consider the model of
non-interacting electrons on a square lattice, with both
nearest-neighbor and next-nearest-neighbor hoppings, in a
uniform magnetic field with rational flux, $\phi=p/q$, per
plaquette. In general, the one-electron spectrum is split
into  bands. If the Fermi level is in a gap between
two  bands, the Hall conductance is quantized. At
some special values for hopping parameters, the gap may
close, leading to a Hall plateau transition. \cite{gap}
One approach to study such transitions is to use
effective field theory. The basic idea is that in the
neighborhood of gap closing points, energy-momentum
dispersion is generally linear, corresponding to the
appearance of massless Dirac fermion(s). Thus the
large-distance behavior of the system near the
transition can be described by an effective field
theory with Dirac fermions, in which the gap closing
and reopening are represented by sign change in fermion
masses. (A similar situation is the anyon Mott transition
near the fermion point, which has been studied in this
way in ref.  \cite{cfw}.) Usually one is tempted to
include into the effective theory only the massless
fermions that appear when the gap closes; it is these
fermions whose Dirac mass may change sign across the
transition. Indeed, it has been generally shown that
counting only these massless Dirac fermions at the
transition point will give us the correct value for
the discrete jump in Hall conductance across the
transition. \cite{oshikawa} Also only the massless Dirac
fermions contribute to the critical exponents for the
transition. \cite{cfw} However, as we will see, this does
not guarantee that the Hall conductance before and after
the transition can be correctly reproduced. In other
words, the topological nature of the Hall conductance
dictates a matching condition for the large-distance
effective field theory to be consistent with the
microscopic lattice formulation.

To see the necessity of such consistency conditions,
we note that the contribution of a planar massive
Dirac fermion to Hall conductance is known to be plus
or minus one-half in units of $e^2/h$, with the sign
depending on the sign of Dirac mass. Therefore, the
change in Hall conductance is always an integer, when
a number of fermion masses change sign. But, if an odd
number of Dirac masses change sign across the transition,
which is possible as we will see below in a numerical
example, the Hall conductance either before or after
the transition is half-odd-integer, in conflict with the
integer quantization as required by the microscopic
topological arguments. Thus a certain number
of massive Dirac fermions, whose masses do not
change sign across the plateau transition, must
be present in the effective field theory. We will
call such fermions as spectators, since really
they do not participate in the phase transition
because they remain massive at the transition point.
However, they are there in the effective theory,
not only ensuring the Hall conductance to be integral,
but also, added to the non-triviality of the problem,
ensuring that the integral value of the Hall
conductance before or after plateau transition
to be the {\it correct one} as dictated by the microscopic
 topological invariant. So the number of species of
these massive ``spectators'' can not be arbitrary; it
needs to be determined from first principles.
Previously the necessity of the existence of massive
``spectator'' fermions has been noticed in Ref.
 \cite{cfw} and  \cite{haldane} in some simple and
specific models for phase transitions to quantum Hall
states. Here we will discuss a general prescription
and a concrete procedure for determining all the
massive spectators in more complicated
models with arbitrary rational flux per plaquette.

The paper is organized as follows. First in the next
section, we briefly review the (2+1)-dimensional
Dirac fermions, the associated zero mode and
resulting contribution, $\pm (1/2) e^2/h$ per species,
to Hall conductance, providing arguments for the
necessity of having massive spectators in the theory.
In Sec. III, we propose a general prescription for
determining the number of massive spectators, i.e.
counting the number of ``vortices'' in the magnetic
Brillouin zone for the highest filled  band
near the transition. A concrete procedure in
which we find vortices by edge states
is introduced.  To exemplify, we use a tight binding
hamiltonian with nearest-neighbor (NN)
and next-nearest-neighbor (NNN)
hoppings to illustrate our procedure. Sec. IV is
devoted to exhibiting numerical results in this model,
showing how our procedure works when gap closing occurs
near certain critical value of the
NNN hopping parameter. The existence of
massive spectators is clearly demonstrated in the
 example. Finally in Sec. V we summarize our
results and discuss relationship of our Hall-conductance
matching condition to precedents in the literature
on matching topological property for theories at
small and large distances.

\section{Dirac fermions in (2+1)-Dimensions}

First let us briefly review the Hall conductance for
a system of planar Dirac fermions in the effective
field theory,  with ``relativistic'' energy-momentum
relation, which becomes linear for the massless case.
This quantity can be extracted from a simple
calculation of the ``vacuum-polarization'' diagram
for the photon propagator with a fermion loop. This
one-loop diagram is superficially divergent, a certain
regularization is needed. Then, the one-loop integral
becomes finite even before removing the cut-off, with
the result  \cite{djt,nsw}
\begin{equation}
\sigma_{xy}=-{\frac 1 2}\, \sum_{\alpha}\,
 {\rm sgn}\, (m_\alpha ).
  \label{eq:dirac_hall}
\end{equation}
Here $m_\alpha$ is the Dirac mass for species $\alpha$.
It has been shown that if the fermions remain massive,
there is no correction to the one-loop value of
$\sigma_{xy}$ from higher orders. \cite{ch,ssw} Other
derivations without computing diagrams can be found
in references \cite{ns,redlich,jackiw}. Here we present an argument
 \cite{haldane,jackiw} which does not involve any diagram
and regularization explicitly.

The hamiltonian for a planar ``relativistic''
fermion of species $\alpha$ in a static,
uniform, perpendicular magnetic field is given by
 \begin{equation}
H_\alpha = \Pi_\alpha^1{\bf \sigma}^2+\Pi_\alpha^2
{\bf \sigma}^1 +m_\alpha{\bf\sigma}^3,
   \label{eq:dirac_ham}
 \end{equation}
where $\Pi_\alpha^1$ and $\Pi_\alpha^2$ are spatial components
of covariant derivative, satisfying $[\Pi_\alpha^1,\Pi_\alpha^2]=
ie\hbar B$, $B$ is the magnetic field strength, and
${\bf \sigma}^{(1,2,3)}$ are Pauli matrices.
The spectrum for a single species at $B=0$ is symmetric
with respect to zero and is given by
\begin{equation}
\epsilon_\alpha^{\pm}(k) =\pm \sqrt{(\hbar k)^2 +m_\alpha^2},
 \label{eq:spec_withoutB}
\end{equation}
which is gapless when $m_\alpha=0$, and is unbounded both
from below and from above.  When $B\neq 0$, relativistic Landau
levels are given by
\begin{eqnarray}
\epsilon_{\alpha,n}^{\pm}&=&\pm \sqrt{n \hbar |e B| +m_\alpha^2},
\,\,\,\, ( n\ge 1)\, \\
\epsilon_{\alpha,0}&=& m_\alpha \, {\rm sgn} ( e B ).
  \label{eq:spec_withB}
\end{eqnarray}

Note that there is a zero-mode (\ref{eq:spec_withB})
 labeled by $n=0$, which
depends on the sign of fermion mass. For the vacuum state,
whether this mode is filled or not depends on the sign of
mass: filled if $m_\alpha<0$, and not if $m_\alpha>0$. When
$m_\alpha=0$, there is a charge conjugation symmetry for this
species.
In the mean time, the vacuum is doubly degenerate, with the
zero-mode filled or not filled. In the continuum field theory,
this ambiguity can not be resolved. It has been argued
\cite{jackiw} that a fermion-number fractionalization
occurs if one assumes charge conjugation symmetry:
the fermion number of the doubly degenerate
vacuum is either $+ 1/2$ or $ -1/2$ depending on
whether the zero mode is filled or empty.
Similarly for the Hall conductance, we may argue that
since the contribution of the filled zero-mode is always
unity (in units of $e^2/h$), charge conjugation
symmetry requires that the vacuum with or without
the zero-mode filled should have fractional contribution
$\pm (1/2) e^2/h$, respectively. Now let us add an
infinitesimal mass to the fermion to remove the vacuum
degeneracy. Obviously for the vacuum state, the zero-mode
is filled if $m_\alpha <0$, and empty if $m_\alpha > 0$.
Thus, we obtain the value of $\sigma_{xy}$ to be $\mp
(1/2) e^2/h$, depending on the sign of mass.
When there are several species of fermions, the Hall
conductance is given by a sum of their individual
contributions, as in Eq.(\ref{eq:dirac_hall}).

Therefore, if the mass of a species changes sign across a
plateau transition, it gives rise to a change in Hall
conductance by unity. So to account for the change in Hall
conductance, one only needs to look for massless fermions
at the transition.

However, it is easy to see that if the number of massless
fermions is odd at the transition, their total contribution to
Hall conductance before or after the transition is a half-odd
integer. If this gives the total Hall conductance, it is in
conflict with integral quantization of Hall conductance
implied by the TKNN topological invariant.
One may wonder perhaps this would never happen, similar to
the no-go theorem of Nielson and Ninomiya \cite{nogo} in
$3+1$-dimensions, which asserts that in the
continuum limit of a lattice theory the number of massless
chiral fermions must be even. If a similar theorem could hold
also in the $2+1$-dimensional case, perhaps the number of
massless Dirac fermions must be even. Unfortunately, it is not
the case: No such no-go theorem exists in $2+1$-dimensions.
Moreover, in Sec. 4 we will show a numerical example, in which
gap closing occurs clearly at three isolated points in the
magnetic Brillouin zone, corresponding to three
(no more and no less) massless fermions.

Thus if the low energy spectrum of the system
consists of unpaired (massless) fermions, there
must be massive fermions present in the ``high-energy''
sector of the theory. These massive fermions remain
massive during the transition and, therefore, do not
participate in the Hall plateau transition,
in the sense that their contribution to Hall
conductance does not change across the transition.
Therefore they are ``spectators'' of the transition.
But their existence restores the integral quantization
of Hall conductance before or after the transition.
\cite{cfw,haldane} In the next section we will discuss
the general principle, as well as a practical procedure,
to determine the number of species of the ``hidden''
massive fermion spectators from microscopic information.

\section{ Vorticities and edge states in the NNN model }

\label{sec:gap_and_dirac}
For definiteness, let us consider a tight-binding
hamiltonian on a square lattice with both NN and NNN
hoppings. In the Landau gauge, the  hamiltonian  is given by
\begin{eqnarray}
H  = & & -t  \sum_{m,n} (
 c_{m+1,n}^\dagger
c_{m,n}
+
c_{m,n+1}^\dagger e^{i 2\pi \phi m}
c_{m,n} ) \nonumber \\
& & -t_d  \sum_{m,n}
c_{m+1,n+1}^\dagger e^{i 2\pi \phi( m + 1 /2 )}
c_{m,n}
 -t_d'  \sum_{m,n}
c_{m,n+1}^\dagger e^{i 2\pi \phi ( m + 1 /2 )}
c_{m+1,n}
+ H.c.
\label{eq:hmlt}
\end{eqnarray}
where $c_{m,n}$ is the annihilation operator for a
lattice fermion at site $(m,n)$.  We assume that
the magnetic flux per plaquette $\phi$ is rational,
{\it i.e.\/}, $\phi=p/q$ with mutually prime integers
$p$ and $q$.

The Hall conductance in this case is a bulk quantity
which is described by the TKNN topological integer
 \cite{tknn}.
It is written as the Chern number of a $U(1)$
fibre bundle over the magnetic Brillouin zone  as
\begin{equation}
        \sigma^{j,\ bulk}_{xy} =- {\frac {e^2} h}
\frac 1 {2\pi i}
\int\int_{T^2_{MBZ}}{dk_x dk_y} \ [ \ {\bf \nabla}_k \times
{\bf A}_u^j({\bf k})\ ]_z,
\label{eq:chern}
\end{equation}
\begin{equation}
{\bf A}_u^j({\bf k}) =
 \langle u^j({\bf k}) |{\bf \nabla}_k |u^j({\bf k})\rangle
=\sum_{m=1}^q
{u^j_m}^*({\bf k}){\bf \nabla}_k u^j_m({\bf k}),
\end{equation}
where $u^j_m$ is a normalized Bloch function of the $j$-th
energy band, with $m$ $(1\leq m \leq q)$ labeling its
components. (See references for precise definitions.
 \cite{mkann,chern_yh})
 An important observation here is that
the magnetic Brillouin zone $T^2_{MBZ}$ is topologically a
torus rather than a rectangle. Since the torus does not
have a boundary, application of Stokes' theorem
to Eq.(\ref{eq:chern}) would give $\sigma_{xy}=0$
if ${\bf A}_u^j({\bf k}) $ is well defined
on the entire torus $T^2_{MBZ}$.
A possible non-zero value of $\sigma_{xy}$
is a consequence of a non-trivial
topology of ${\bf A}_u^j({\bf k}) $.
In order to better understand the
relevant topology, let us examine the
effect of a phase transformation
of the wavefunction
\begin{equation}
|u^j({\bf k})\rangle ' =
e^{i f({\bf k })}  |u^j({\bf k})\rangle,
\label{eq:gt1}
\end{equation}
where $f({\bf k })$ is an arbitrary smooth
function of $\bf k$ over the Brillouin zone.
The corresponding gauge transformatin for
${\bf A}_u^j({\bf k}) $ is
\begin{equation}
{\bf A}_u^j({\bf k})' ={\bf A}_u^j({\bf k})
+ i \nabla _k f({\bf k }),
\label{eq:gt2}
\end{equation}
which clearly leaves $\sigma_{xy}$  invariant.
The non-trivial topology arises when the phase
of the wavefunction can not be determined uniquely
and smoothly in the entire Brillouin zone.
The gauge transformation defined above implies
that the overall phase of the wavefunction can be
chosen arbitrarily. It can be determined, for example,
by demanding the $q$-th component of
$|u^j({\bf k})\rangle$ to be real.
However, this is not enough to fix the
phase over the Brillouin zone, when $u^j_q({\bf k})$
vanishes at some points. Let us denote these
zeros by $k^{*}_{s}$ with $s=1,\cdots,N $,
and define small regions around the zeros
by $ R_s^\epsilon = \{ \ { \bf k} \in T^2_{MBZ} \Bigl|
\ | {\bf k}-{\bf k}^*_s|<  \epsilon,\
\Psi^j_q({\bf k}_s^*)=0 \}$,  $\epsilon>0$.
We may adopt a different phase convention in
$R_s $ so that another component, say,
$u^j_1({\bf k})$, is real.
(We denote it by $|v^j({\bf k })\rangle$).
Then the overall phase is uniquely determined
on the entire Brillouin zone $T^2_{MBZ}$.
At the boundaries, $\partial R_s$, we have
a phase mismatch
\begin{equation}
|v^j({\bf k})\rangle =
e^{i f({\bf k })}  |u^j({\bf k })\rangle.
\end{equation}
By using the above formulas for gauge
transformation, Eqs. (\ref{eq:gt1}),
(\ref{eq:gt2}), we have
\begin{eqnarray}
\sigma^j_{xy}
&= &
- \sum_{s=1}^{N} n_s\, , \\
n_s&=& \frac 1 {2\pi }
\oint_{\partial R_s}
 {\bf \nabla}f ({\bf k}).
\label{eq:oint}
\end{eqnarray}
Here $n_s$ must be integers since each of the states
vectors must fit togther exactly when we complete
full revolutions around each $R_s$.
This implies that the zeros of a certain component
of the Bloch function define vortices in the
Brillouin zone, whose integral vorticities
contribute to the Hall conductance.
While the phase of the wavefunction depends
on phase convention (gauge choice),
the total vorticity is a gauge invariant quantity.
In this way, in principle, counting the total
vorticity of the $U(1)$ phase of the Bloch wavefunction
gives the bulk Hall conductance. But this would
need the knowledge of (explicit or numerical)
wavefunctions in the whole Brillouin zone.

A more practical prescription to obtain
the vortices of the bulk states is proposed
in Refs. \cite{edge_yh} and \cite{chern_yh},
which are summarized as follows. This method
exploits a relationship between two closely
related systems. We use periodic boundary
conditions in $y$-direction. But in
$x$-direction, we consider two possible
boundary conditions separately: the periodic
and fixed ones. Using the periodic boundary
conditions in both directions
and taking the infinite size limit, one can
obtain a bulk system (without edges).
By using the fixed boundary conditions
in $x$-direction, one can consider a
cylindrical system with edges in
that direction. In this case,
by a Fourier transformation in
$y$-direction, we obtain a
one-dimensional tight-binding equation
with parameter $k_y$,
\begin{eqnarray}
- ( t & + & t_d e^{i\Lambda(m)}
+   t_d' e^{-i\Lambda(m)}) \Psi_{m+1}(k_y)
- 2 t \cos (-k_y + 2\pi \phi m) \Psi_{m}(k_y)
\nonumber \\
&-& ( t + t_d'e^{i\Lambda(m-1)}
+ t_d e^{-i\Lambda(m-1)})\Psi_{m-1}(k_y)
 = E(k_y) \Psi_{m}(k_y),
\label{eq:1dhmlt}
\end{eqnarray}
where
\begin{equation}
\Lambda(m)= -k_y+2 \pi \phi m +\pi\phi,
\ \ \ \Lambda(m+q)=\Lambda(m) .
\label{eq:lambda}
\end{equation}
This difference equation can be written as

\begin{equation}
\left( \begin{array}{c}
           \Psi_{m+1}\\
           \Psi_{m}
        \end{array}
\right)
=
{M}_m(E,k_y)
\left( \begin{array}{c}
           \Psi_{m}\\
           \Psi_{m-1}        \end{array}
\right),
\label{eq:tr}
\end{equation}
where $M_{m}$ is the following $2\times 2$ transfer matrix
\begin{eqnarray}
M_m(E,k_y)  & =  &
\left( \begin{array}{c c}
\frac
{-E -  2t \cos (-k_y+2\pi\phi m) }
{t + t_d e^{i\Lambda(m)}  + t_d' e^{-i\Lambda(m)} }
&    -{ \frac
{t + t_d' e^{i\Lambda(m-1)}  + t_d  e^{-i\Lambda(m-1)}}
{t + t_d  e^{i\Lambda(m)}  + t_d' e^{-i\Lambda(m)}}
}   \\
           1    &   0
        \end{array}
\right).
\label{eq:mtramat}
\end{eqnarray}
Then the spectrum is completely governed
by a product of $q$ transfer matrices,
$M(E,k_y)$, given by
\begin{equation}
 M(E,k_y) =
 \prod_{m=1}^q M_m(E,k_y), \ \ \
(| \det  M(\epsilon,k_y) | =1).
\label{eq:trmat}
\end{equation}
In the following,  for simplicity, we assume
$t=1$, $t_d'=t_d$ and $L_x$ to be a multiple of $q$.
The boundary conditions are
$\Psi_0=\Psi_{L_x}=0$
and we set $\Psi_1=1$
to fix the normalization.

The one-dimensional system (\ref{eq:1dhmlt})
has a period $q$. Thus as $L_x\to\infty$,
part of the spectrum converges to $q$ energy
bands which are determined by the condition
\begin{equation}
| {\rm Tr}   M(E,k_y)  | \le 2.
  \label{eq:cnd_band}
\end{equation}
This corresponds to the on-shell condition in
usual scattering theory.
Further, for the cylindrical system,
there are additional eigen energies,
$\mu_j$ ($j=1,\cdots,q-1$), satisfying
\begin{equation}
M_{12}(\mu_j, k_y)=0,
\end{equation}
so that $\Psi_{L_{x}}=0$. For each $\mu_{j}$,
the condition (\ref{eq:cnd_band}) is not satisfied.
On the other hand, the corresponding state
is localized near either $x\approx 0$ or
$x\approx L_x$, depending on the value of
$|M_{22}(\mu_j,k_y)|$:
\begin{eqnarray}
|M_{22}(\mu_j, k_y)| < 1 &:
\ \  {\rm localized\ at\ the \ left\
edge} \nonumber \\
|M_{22}(\mu_j, k_y)| > 1 &:
\ \  {\rm localized\ at\ the \ right\ edge}.
\label{eq:edgwhich}
\end{eqnarray}
So this is an edge state, whose energy lies in
the $j$-th gap. Note that for such state,
the above equation implies
\begin{equation}
\Psi_{q}(\mu_j (k_{y}),\, k_y)= 0\, .
  \label{eq:edge_cond}
\end{equation}
On the other hand, if $|M_{22}(\mu_j(k_{y}), k_y)|=1$,
the condition (\ref{eq:cnd_band})
is satisfied, implying that the edge state
is in touch with an energy band edge. Then the
above equation (\ref{eq:edge_cond}) tells us that
the corresponding $k_{y}$, together with $k_{x}
=0$ or $2\pi/q$ depending on which (top or bottom)
egde it is of the energy band, gives a zero
of the Bloch function $\Psi_{q}$ in the Brillouin zone.
In this way, each time when the edge state becomes  degenerate
with a bulk state at band edge occurs,
we find a zero of the $q$-th component of the
Bloch wave function for that band, which
gives a desired $U(1)$ vortex in Eq. (\ref{eq:chern}).

How to determine the vorticity associated with
each vortex found in this way?
A useful and practical rule is proved in Ref.
\cite{chern_yh}:  When the degeneracy occurs
at the top of a band and the edge state
jumps from the right edge to the left
(or the opposite way from the left edge to the right)
with increasing $k_y$, the vorticity is $+1$ (or
$-1$). Similarly, when the degeneracy occurs
at the bottom of a band, the vorticity is just
opposite to the previous case. In Fig~\ref{f:ex},
a typical example of the degeneracy between
an energy band and edge state is shown.
The energy of an edge state that is localized
at the right edge is drawn by a  solid line and
the one at the left edge by a dotted line.

Since there is one and only one edge state in
the $j$-th gap, we may trace its energy
$\mu_{j} (k_{y})$ as $k_{y}$ vareis from $0$
to $2\pi$, and define a winding number (around
the $j$-th gap) associated with the edge state
by adding all the vorticities at its
degeneracy points. (Geometrically indeed
this number can be realized \cite{chern_yh} as
the winding number of the loop traced
by $\mu_{j} (k_{y})$ around the $j$-th hole
on the complex energy Riemann surface for
Eq. (\ref{eq:1dhmlt}).) It is amusing that
this winding number is exactly the contribution
of the $j$-th band to the Hall conductance.
In this way, a direct relationship between the
bulk Hall conductance and the behavior of the edge
state is established. We will exploit this
relation to find the vorticities of the bulk Bloch
states. As will be discussed in the next section,
our rule is simply {\em to assign a Dirac fermion
to each vortex associated with the edge state} in
the gap relevant to the plateau transition.

\section{Dirac fermions and spectators on the lattice}

In the model with only the NN hopping,
it is known that there are zero modes
when $q$ is even \cite{gap_k,gap,wz}.
There are $q$
zero modes and the dispersion is linear
in momentum near these zero modes.
As far as the low energy physics is concerned,
when the fermi energy is near   zero,
we may treat the system as that of   Dirac fermions.
The energy dispersion for $\phi=1/4$ case is shown
in Fig~\ref{f:Dirac}(a).
There are four energy bands and
the two of them near $E=0$
are degenerate where the linear dispersion
is clearly observed.

Dirac fermions also appear in the NNN
model \cite{gap,oshikawa}.
For example, in Fig~\ref{f:Dirac}(b),
the spectrum for $\phi=1/3$ and $t_d=t_{Dirac}$,
$t_{Dirac}\approx 0.2679t$ is shown.
In this case, there are three energy bands
and the two higher energy bands are
degenerate at three momenta, corresponding
to three massless Dirac fermions.

In general, there are $q$ gap closing points
in the NNN model with $t_d=t_{Dirac}$.
Deviation of $t_d$ from $t_{Dirac}$ brings
a mass term in the energy dispersion.
The formula (\ref{eq:dirac_hall}) tells
that one Dirac fermion carries the Hall
conductance $\pm 1/2 (e^2/h)$ depending on
the sign of its mass. For an even number
of  Dirac fermions  (as shown
in  Fig~\ref{f:Dirac}(a)),
 Eq.(\ref{eq:dirac_hall}) always gives
 an integer $\sigma_{xy}$.
On the other hand, if $q$ is odd,
according to Eq.(\ref{eq:dirac_hall}),
these $q$ Dirac fermions associated
with the gap closing points will give
a half-integer value for any kind of
combination of signs in the mass terms.
It would break the integral quantization
for the Hall conductance of the non-interacting
fermions, if no massive Dirac fermions are
introduced. Indeed, as will be shown below,
for example for the case of
Fig~\ref{f:Dirac}(b)), there is in addition
a massive Dirac fermion, corresponding to
a vortex not passing through any of the
gap closing points.  Though it is a "spectator"
for this Hall transition. it helps
restore the integeral Hall conductance.

In Fig.~\ref{f:1o3},
 the energy bands, edge states, and the corresponding
vortices of the NN model are shown
 where the white circles denote $+1$ vorticities and
the black ones  denote $-1$ vorticities.
In Figs.~\ref{f:edges1o3},
the same things are shown
for the NNN model for several $t_d$'s.
One can see how the second energy gap
closes and then opens again.
The energy dispersions near the gap closing
are shown in Fig.~\ref{f:edges1o3} (d)
for $t_d=0.267945t$ and Fig.~\ref{f:edges1o3}
for (e) $t_d=0.267955t$,
where there are three almost linear
dispersions in the second energy gap.
The exact linear dispersion is realized
at $t_d=t_{Dirac}\approx  0.26795$.
Thus the Dirac Fermion are massive in the cases of
 Figs.~\ref{f:edges1o3} (d) and (e).
Further by comparing Fig.~\ref{f:edges1o3}(c)
and Fig.~\ref{f:edges1o3}(d) \cite{gap},
one can see that the sign of the mass is
reversed after the gap reopening.
Near the gap closing (appearance of the
massless Dirac fermions)
at the three different momentum points,
there are three vortices as shown in
the Figs~\ref{f:edges1o3}.
The vortices moves between the lower
band and the higher band during the process.
It explains the change in the Hall conductance.

What is crucial to our point is that there is
an additional vortex near the $k_y=\pi$ point,
where there is a large energy gap. Though
the shape of the energy dispersion
is quite different from that of a Dirac fermion,
it does not prevent us from including the effect
of this vortex as a Dirac fermion with a
large mass term in the effective field theory,
since its only role in the theory for the Hall
transition is to reproduce the correct
value of the Hall conductance\cite{comm}.
This is the hidden ``spectator'' which restores
the integral quantization of Hall conductance.
In short, it is the existence of the vortices
rather than their energy dispersion that
defines all relevant Dirac fermions appearing
in the effective theory, including massive
spectators.

Note that when one changes the parameters
of the model, the Dirac fermions associated
with vortices in the magnetic Brillouin
zone are always created in pairs. For example,
in Figs.~\ref{f:edges1o3}(a) and (b),
between $t_d=0.24$ and $t_d=0.25003$,
a pair of  Dirac fermions
are created near $k_y=\pi$.
Two fermions move toward the gap closing points
at $k_y=2\pi/3$ and $k_y=4\pi/3$  and there
they go across the energy gap when the gap
closes (Figs.~\ref{f:edges1o3}(a)-(g)). In
summary, for the effective theory of this
quantum Hall transition, we should have
three massless Dirac fermions and one massive
spectator with positive mass.

\section{ Summary  and Discussion}

We have shown that in general one needs to
add massive fermions (spectators)
into the effective field theory in orde to
get correct integral Hall conductance.
The number of spectators is not arbitrary, since the
Hall conductance is dictated by the microscopic
TKNN topological invariant. Therefore,
one needs to consider the ``high-energy'' sector
or the microscopic details of the system,
in order that the Hall conductance
in the low-energy effective field theory
at large distances should be the same
as in the high energy microscopic
model at small distances.
This  matching condition, required by topological
considerations, establishes a connection between
theories of the same system at large and small
distaces. In the models with noninteracting
electrons on a planar lattice, we have
given the principles and a
practical procedure for determining the number of
massive species of fermion spectators from microscopic
information. Numerical examples are shown for the
validity of our prescription and for the
implementation of our practical procedure.

To conclude this paper, we would like to point out the
relationship of our present problem to its precedents
in the literature on topological investigations in
quantum field theory.  From the point of view of the
latter, the Hall effect in a planar system (or in 2+1
dimensions) can be  expressed by a Chern-Simons
effective action, obtained by integrating out the
electrons in the system, for the external
electromagnetic field:
\begin{equation}
S_{eff} (A)= {\sigma_{xy}\over 2}
\epsilon^{\mu\nu\lambda} A_{\mu} F_{\nu\lambda},
\label{eq:CS}
\end{equation}
whose variation with respect to $A_{\mu}$ gives us
the (Hall) current in the system induced by the
external field.  (Here $\mu,\nu,\lambda=0,1,2$.)
There is a well-known formal relationship \cite{ns,zwz}
between this action and famous axial anomaly equation
in 3+1 dimensions:
\begin{equation}
\partial_{\mu} J_{5}^{\mu} = \frac{1}{16\pi^{2}}
\epsilon^{\mu\nu\lambda\sigma} F_{\mu\nu}
F_{\lambda\sigma}.
\label{eq:anomaly}
\end{equation}
Here non-conservation of the axial current,
$J_{5}^{\mu}$, is expressed in terms of
the gauge field that couples to the vector
current. One observes that the right side of
Eq. (\ref{eq:CS}) can be viewed,
up to numerical factors, as the surface term
of an integral of the right side of Eq. (\ref{eq:anomaly}).
It is not surprising to find similarities between
our present problem and some known examples related to
the axial anomaly in 3+1 dimensions.

The first example we like to mention
is the t'Hooft's anomaly-matching condition. \cite{tHooft}
In 3+1 dimensions, chiral fermions give rise to
axial anomaly, which is known to be of topological origin.
If this anomaly exists at the fundamental
constituent level, t'Hooft has argued
that it must survive at large distances even in theories
in which the massless fermions are confined (like quarks)
and do not show up at large distances. This requirement,
i.e. the axial anomaly in the effective theory at large
distances should match that in the fundamental theory at
small distances, imposes constraints on the spectrum of
the effective theory at large distances. In certain cases,
massive ``spectators'' may be needed to satisfy this
condition. In our problem, what plays the role of
axial anomaly is the Hall conductance in 2+1 dimensions.
The lattice theory we start with is the miscroscopic theory
 and  the effective field theory is the one at large distances.

In certain sense, the Nielson-Ninomiya ``no-go'' theorem
for unpaired massless chiral fermions in the continuum
limit of a lattice theory constitutes another counterpart
of our problem in 3+1 dimensions. The basic argument for
this ``no-go'' theorem is again the necessity of
matching, in the continuum limit, the vanishing axial
anomaly in the lattice theory.

Therefore, our present problem can be viewed as a
2+1 dimensional version of the above two problems
for matching topological properties for theories of
the same system at small and large distances.

\section{Acknowledgments}
Y.H, M.K, and Y.S.W were supported in part by  Grant-in-Aid
from the Ministry of Education, Science and Culture
of Japan.
Y.S.W. thanks the Institute for
Solid State Physics, University of Tokyo for warm hospitality
during his visit.
 His work was supported in part by US NSF grant No. PHY-9308458.

\begin{figure}
\caption{ Typical example for the degeneracy between
the energy band and the edge states. It gives the vortex.
 \label{f:ex}}
\end{figure}

\begin{figure}
\caption{ Energy spectrum of the two-dimensional
tight binding model:
(a)$\phi=1/4$ and $t_d=0.0$,
(b)$\phi=1/3$ and $t_d=t_{Dirac},t_{Dirac}\approx 0.2679$.
 \label{f:Dirac}}
\end{figure}

\begin{figure}
\caption{ Energy spectrum of the two-dimensional
tight-binding model with  $t_d=0$ and $\phi=1/3$ (NN model).
The energy of the edge states
and the corresponding vortices
on the energy bands are also shown.
The solid line is the edge state localized at the right edge
and the dotted line is one localized at the left edge.
The vorticities are denoted by the white or black circles.
The black one is vorticity $+1$ and the white one is $-1$.
 \label{f:1o3}}
\end{figure}

\begin{figure}
\caption{ Energy spectrum of the two-dimensional
tight-binding model with next nearest neighbor hopping $t_d$ (NNN model).
The energy of the edge states and the corresponding vortices on the energy
bands are
also shown for $\phi=1/3$:
(a) $t_d=0.24$,(b) $t_d=0.25003$, (c) $t_d=0.263$,
 (d) $t_d=0.267945$,
 (e) $t_d=0.267955$,
(f) $t_d=0.273$ and (g)
$t_d=0.4$.
 \label{f:edges1o3}}
\end{figure}


\begin{references}
\bibitem{comm}
In principle,
we may include some additional parameters in the hamiltonian and
change the energy dispersion near $k_y=\pi$ adiabatically to a
form similar to that of the Dirac fermion. During the process,
the energy gap remains finite and thus the Hall conductance
remains the same.
\end{references}
\end{document}